\documentstyle[12pt,titlepage]{article}

\font\grande=cmr9.5 scaled \magstep4
\font\medio=cmr9.5 scaled \magstep2
\outer\def\beginsection#1\par{\medbreak\bigskip
      \message{#1}\leftline{\bf#1}\nobreak\medskip
\vskip-\parskip
      \noindent}

\def\laq{\raise 0.4ex\hbox{$<$}\kern -0.8em\lower 0.62
ex\hbox{$\sim$}}
\def\gaq{\raise 0.4ex\hbox{$>$}\kern -0.7em\lower 0.62
ex\hbox{$\sim$}}

\begin{document}
\bibliographystyle {unsrt}

\titlepage

\begin{flushright}
CERN-PH-TH/2006-055
\end{flushright}

\vspace{15mm}
\begin{center}
{\grande Transfer matrices for magnetized CMB anisotropies}\\
\vspace{15mm}
 Massimo Giovannini 
 \footnote{Electronic address: massimo.giovannini@cern.ch} \\
\vspace{6mm}

\vspace{0.3cm}
{{\sl Centro ``Enrico Fermi", Compendio del Viminale, Via 
Panisperna 89/A, 00184 Rome, Italy}}\\
\vspace{0.3cm}
{{\sl Department of Physics, Theory Division, CERN, 1211 Geneva 23, Switzerland}}
\vspace*{2cm}

\end{center}

\vskip 2cm
\centerline{\medio  Abstract}
Large-scale magnetic fields can affect scalar 
cosmological perturbations whose evolution is 
described in the conformally Newtonian gauge and within 
the tight coupling approximation.  The magnetized curvature perturbations present after matter radiation equality (and prior to decoupling) are computed in terms of an appropriate transfer matrix allowing a general estimate of the Sachs-Wolfe plateau. From the observation that CMB initial conditions should be (predominantly) adiabatic, the contribution of the magnetic field intensity can be constrained.
\noindent

\vspace{5mm}

\vfill
\newpage
Large-scale magnetic fields are observed at a $\mu$ G level in galaxies, clusters and in some superclusters \cite{f1}. 
Compressional amplification (taking place during the gravitational collapse of the protogalaxy) allows to connect the observed magnetic field to a protogalactic field, present prior to gravitational collapse, of typical strength of $0.1$ nG.  
A better understanding of the interplay between dynamo  theory 
and the global conservation laws of magnetized plasmas
 has been recently achieved \cite{f2} also because of the improved 
 comprehension of the solar dynamo action. 
 It is then plausible, within the dynamo hypothesis,  that 
the protogalactic field could be even much smaller than the nG and still explain some crucial properties of our magnetized Universe.

Thanks to magnetic flux (and magnetic helicity) conservation 
in a conductive plasma, a magnetic field of nG strength at the epoch of galaxy formation can be as large as ${\rm mG}$ (i.e. roughly 6 orders of magnitude larger) at the epoch of photon decoupling, i.e. for $z_{\rm dec} = 1100$.
If large-scale magnetic fields have primordial origin, 
they were present prior to matter-radiation equality affecting, potentially,  
CMB anisotropies \cite{f1}. Through the years, various studies have been 
devoted to the effect of large-scale magnetic fields on the vector and tensor CMB anisotropies \cite{f3} (see also \cite{f4} and references therein for some recent review articles). 

The implications of fully inhomogeneous magnetic fields on the scalar modes of the geometry remain comparatively less explored. 
By fully inhomogeneous we mean stochastically distributed fields 
that do not break the spatial isotropy of the background \cite{f4}.
One of the aims of the present paper is to partially bridge this gap and to
open the way for further developments.  In short the idea is the following.
The simplest set of initial conditions for 
CMB anisotropies, implies, in a $\Lambda$CDM framework, that a nearly 
scale-invariant spectrum of adiabatic fluctuations is present after 
matter-radiation equality but before decoupling for typical wavelengths 
larger than the Hubble radius at the corresponding epoch \cite{f5}. 
It became relevant, through the years, to relax the assumption
of exact adiabaticity and to scrutinize the implications of a more general
mixture of adiabatic and non-adiabatic initial conditions (see \cite{f6} and references therein). In this paper it will be argued, along a similar perspective, that large-scale magnetic fields slightly modify the adiabatic paradigm so that their typical strengths may be constrained. To achieve such a goal, the first step is to 
solve the evolution equations of magnetized cosmological perturbations 
well before equality. The second step is to follow the solution through 
equality (and up to decoupling). On a more technical ground, the second step amounts to the calculation of the so-called transfer matrix whose specific form is one of the the subjects of the present analysis.

Consider then the system of cosmological perturbations 
of a flat Friedmann-Robertson-Walker (FRW) Universe, 
characterized by a conformal time scale factor $a(\tau)$,  
and consisting of a mixture of photons, baryons, CDM particles and massless neutrinos. In the conformally Newtonian gauge \cite{f6,f7} the scalar fluctuations 
of the metric tensor $g_{\mu\nu}$ are parametrized in terms of 
 the two longitudinal fluctuations i.e. $\delta g_{00} = 2 a^2 \phi(\tau,\vec{x})$ and $\delta g_{ij} = 2 a^2 \psi(\tau,\vec{x}) \delta_{ij}$. 
 The Hamiltonian and momentum constraints, stemming from the $(00)$ and $(0i)$ 
 components of the perturbed Einstein equations are:
\begin{eqnarray}
&& \nabla^2 \psi - 3 {\cal H} ( {\cal H}\phi + \psi') = 4\pi G a^2 [ \delta \rho_{\rm t} + 
\delta\rho_{\rm B}],\qquad \delta \rho_{\rm B}(\tau,\vec{x}) = \frac{B^2(\vec{x})}{8\pi a^4(\tau)},
\label{Ham1}\\
&&\nabla^2( {\cal H} \phi + \psi') = - 4\pi G a^2 (p_{\rm t} + \rho_{\rm t}) 
\theta_{\rm t},
\label{Mom1}
\end{eqnarray}
where ${\cal H} = a'/a$ and the prime denotes a derivation with respect 
to the conformal time coordinate $\tau$. The total energy and pressure 
densities of the mixture, i.e.  $\rho_{\rm t} = \sum_{\rm a} \rho_{\rm a}$ and $p_{\rm t} = \sum_{\rm a} p_{\rm a}$, determine the evolution of the background
geometry according to Friedmann equations: 
\begin{equation}
{\cal H}^2 = \frac{8\pi G}{3} a^2 \rho_{\rm t},\qquad 
{\cal H}^2 - {\cal H}' = 4\pi G a^2 (\rho_{\rm t} + p_{\rm t}),\qquad 
\rho_{\rm t}' + 3 {\cal H} (\rho_{\rm t} + p_{\rm t})=0.
\label{FL}
\end{equation}
In Eqs. (\ref{Ham1}) and (\ref{Mom1}) 
$\delta \rho_{\rm t}$ and $\theta_{\rm t}$, denote, respectively, the total 
density fluctuation of the fluid mixture and the divergence of the total velocity field (i.e. $\theta_{\rm t}
= \partial_{i} v^{i}_{\rm t}$) whose expressions, in terms of the four 
components of the plasma, i.e. $\nu$, $\gamma$, ${\rm c}$ (CDM) and ${\rm b}$ (baryons), is 
\begin{equation}
\delta\rho_{\rm t} = \sum_{{\rm a}} \delta\rho_{\rm a},\qquad
\delta p_{\rm t} = \sum_{{\rm a}} \delta p_{\rm a},\qquad
(p_{\rm t} + \rho_{\rm t}) \theta_{\rm t} = \sum_{{\rm a}} (p_{\rm a} + \rho_{\rm a})
\theta_{\rm a}.
\label{sumdef}
\end{equation}
The spatial components of the perturbed Einstein equations, imply, instead 
\begin{eqnarray}
&& \psi'' + {\cal H} ( \phi' + 2 \psi') + ( 2 {\cal H}' + {\cal H}^2) \phi + 
\frac{1}{3} \nabla^2(\phi - \psi) = 4\pi G a^2 (\delta p_{\rm t} + \delta p_{\rm B}),
\label{tij}\\
&& \nabla^4 ( \phi - \psi) = 12 \pi G a^2 [ 
(p_{\nu} + \rho_{\nu}) \nabla^2 \sigma_{\nu} +
 (p_{\gamma} + \rho_{\gamma}) \nabla^2 \sigma_{\rm B}],\qquad 
 \delta p_{\rm B} = \frac{\delta \rho_{\rm B}}{3}.
 \label{anis1}
 \end{eqnarray}
In Eq. (\ref{anis1}) $\nabla^2 \sigma_{\nu}$ is the neutrino anisotropic stress, while  $\nabla^2 \sigma_{\rm B}$ is the magnetic field anisotropic 
stress defined as:
 \begin{equation}
 \nabla^2 \sigma_{\rm B} = 
 \frac{3}{16\pi a^4 \rho_{\gamma}} \vec{\nabla}\cdot [
  (\vec{\nabla}\times \vec{B})
 \times \vec{B}] + 
 \frac{\nabla^2 \Omega_{\rm B}}{4},\qquad \Omega_{\rm B}(\vec{x}) = \frac{\delta\rho_{\rm B}(\tau, \vec{x})}{\rho_{\gamma}(\tau)},
 \label{magndef}
 \end{equation}
 where, $\Omega_{\rm B}(\vec{x})$ is the magnetic energy density referred to the photon energy density and it is constant to a very good approximation if magnetic flux is frozen into the plasma element 
 \cite{f1,f3,f4}, as assumed throughout the paper. 
The  induced Ohmic current $\vec{J}$ is solenoidal 
(in the magnetohydrodynamical (MHD) description adopted here) and it 
is simply given by $ 4\pi \vec{J} = \vec{\nabla}\times \vec{B}$.  Moreover, in MHD, 
$\vec{E} = \vec{J}/\sigma \simeq (\vec{\nabla}\times \vec{B})/\sigma$ (where
$\sigma$ is the conductivity). Since, prior to decoupling,  
 the Universe was a rather good conductor \cite{f4},
the contribution of the electric energy density and of the  Poynting vector  appearing, in principle, in  Eqs. (\ref{Ham1}) and (\ref{Mom1}) can be safely 
neglected for typical length scales much larger than the screening length 
of the plasma.
It should be stressed that, in Eq. (\ref{anis1}), on top of the 
magnetic piece, the only contribution to the anisotropic stress of the fluid mixture comes from massless neutrinos\footnote{If neutrinos would have a mass in the meV range, they would be non-relativistic today but they will still be counted as radiation prior to decoupling.}(that are collisionless for temperatures smaller than $1$ MeV) and it is parametrized by $\sigma_{\nu}$. The evolution of 
$\delta \rho_{\rm t}$ can be determined from the covariant conservation of the (total) energy-momentum tensor:
 \begin{equation}
\delta\rho_{\rm t}' - 3 \psi' (p_{\rm t} + \rho_{\rm t}) + (p_{\rm t} + \rho_{\rm t}) \theta_{\rm t} + 3 {\cal H}( 1 + c_{\rm s}^2 ) \delta \rho_{\rm t} + 3 {\cal H} \delta p_{\rm nad} = \frac{\vec{E}\cdot \vec{J}}{a^4}, 
 \label{DC}
\end{equation}
where $c_{\rm s}^2 = p_{\rm t}'/\rho_{\rm t}'$ is the (total) sound speed and where 
the (total) pressure density fluctuation $\delta p_{\rm t}$ has been slpit into the adiabatic contribution (i.e. $c_{\rm s}^2 \delta \rho_{\rm t}$) supplemented 
by the non-adiabatic pressure density fluctuation (i.e. $\delta p_{\rm nad}$). The electromagnetic contribution appearing in  Eq. (\ref{DC}) 
contains an electric field and it is therefore suppressed. 

The evolution of the CDM component feels indirectly the presence 
of the magnetic field intensity through the Hamiltonian constraint 
(\ref{Ham1})  and the relevant equations are 
\begin{equation}
\theta_{\rm c}' + {\cal H} \theta_{\rm c} + \nabla^2 \phi=0,\qquad 
\delta_{\rm c}' = 3 \psi' - \theta_{\rm c}, \qquad \delta_{\rm c} = \frac{\delta\rho_{\rm c}}{\rho_{\rm c}}.
\label{CDMeq}
\end{equation}
The neutrinos are coupled to the magnetic field through the Hamiltonian 
constraint (\ref{Ham1}) and through Eq. (\ref{anis1}) (involving the neutrino
anisotropic stress $\nabla^2 \sigma_{\nu}$):
\begin{equation}
\theta_{\nu}'  +\frac{1}{4} \nabla^2 \delta_{\nu} + \nabla^2 \phi = \nabla^2 \sigma_{\nu},\qquad \delta_{\nu}'= 4\psi' - \frac{4}{3} \theta_{\nu}, \qquad 
\sigma_{\nu}' = \frac{4}{15} \theta_{\nu},
\label{nueq}
\end{equation}
where, in full analogy with Eq. (\ref{CDMeq}) the neutrino density 
contrast $\delta_{\nu}$ has been introduced.
Photons and baryons are tightly coupled by Thompson scattering and form, effectively, a single fluid characterized by a velocity field $\theta_{\gamma{\rm b}} = \theta_{\gamma} = \theta _{\rm b}$. The relevant evolution equations are, in this case, 
\begin{eqnarray}
&& \delta_{\gamma}' = 4\psi' - \frac{4}{3} \theta_{\gamma{\rm b}}, \qquad 
 \delta_{\rm b}' = 3 \psi' - \theta_{\gamma{\rm b}},\qquad R_{\rm b}(\tau)) = \frac{3}{4} \frac{\rho_{\rm b}(\tau)}{\rho_{\gamma}(\tau)} = 
\biggl( \frac{698}{z + 1}\biggr) \biggl( \frac{h^2 \overline{\Omega}_{\rm b}}{ 0.023}\biggr),
 \label{pb1}\\
&& \theta_{\gamma{\rm b}}' + \frac{{\cal H} R_{\rm b}}{(1 + R_{\rm b})} \theta_{\gamma{\rm b}} + \frac{\nabla^2 \delta_{\gamma}}{4 ( 1 + R_{\rm b})} + 
\nabla^2 \phi = \frac{3}{4} \frac{\vec{\nabla}\cdot[ \vec{J} \times \vec{B}]}{a^4 
\rho_{\gamma} ( 1 + R_{\rm b})},
\label{pb2}
\end{eqnarray}
where $R_{\rm b}$ is the baryon to photon ratio that depends on the redshift 
$z$. Deep in the radiation-dominated epoch i.e. for $\tau \ll \tau_{\rm eq}$ the 
solution for the magnetized adiabatic mode can be obtained, in Fourier space,  by solving, simultaneously, Eqs. (\ref{CDMeq}), (\ref{nueq}) and (\ref{pb1})--(\ref{pb2}). 
The compatibility of the magnetized adiabatic mode  with 
Eqs. (\ref{Ham1})--(\ref{Mom1}) and also with Eqs. (\ref{tij})-(\ref{anis1})
fixes the integration constants. Defining as $k$ the Fourier (comoving) wave-number we shall be interested in  wavelengths much larger than the 
Hubble radius, i.e. $k \tau < 1$. For $\tau \ll \tau_{\rm eq}$, the 
density contrasts for the magnetized adiabatic mode are, to lowest order 
in $k\tau<1$,
\begin{equation}
\delta_{\gamma} = \delta_{\nu} = - 2\phi_{\rm i} - R_{\gamma} \Omega_{\rm B}, \qquad 
\delta_{\rm b} = \delta_{\rm c} = - \frac{3}{2} \phi_{\rm i} 
- \frac{3}{4} R_{\gamma} \Omega_{\rm B},
\label{DCad}
\end{equation}
where the fractional contribution of photons to the radiation plasma, i.e. $R_{\gamma}$ has been introduced and it is related to $R_{\nu}$, i.e. 
the fractional contribution of massless neutrinos, as
 \begin{equation}
R_{\gamma} = 1 - R_{\nu}, \qquad R_{\nu} = \frac{r}{1 + r},\qquad r= \frac{7}{8} N_{\nu} \biggl(\frac{4}{11}\biggr)^{4/3} \equiv  0.681 \biggl(\frac{N_{\nu}}{3}\biggr).
\end{equation}
From Eqs. (\ref{CDMeq}), (\ref{nueq}) and (\ref{pb2}) the  velocity fields of the various species are
\begin{equation}
\theta_{\gamma{\rm b}} = \frac{k^2 \tau}{4} [ 2 \phi_{\rm i} + R_{\nu} \Omega_{\rm B} - 4 \sigma_{\rm B} ], \qquad \theta_{\rm c} = \frac{k^2 \tau}{2}\phi_{\rm i},
\qquad \theta_{\nu} = \frac{k^2 \tau}{2}\biggl[ \phi_{\rm i}- \frac{R_{\gamma} \Omega_{\rm B}}{2} \biggr]  + k^2 \tau \frac{R_{\gamma}}{R_{\nu}} \sigma_{\rm B}.
\label{VF}
\end{equation}
The quantities $\psi_{\rm i}(k)$ and $\phi_{\rm i}(k)$ appearing in Eqs. (\ref{DCad}) and (\ref{VF})   denote the super-Hubble fluctuations 
that are initially present prior to equality. By solving in terms of the neutrino
anisotropic stress $\sigma_{\nu}$ and by recalling Eq. (\ref{anis1}) the relation 
between $\psi_{\rm i}$  and $\phi_{\rm i}$ can be obtained:
\begin{equation}
\psi_{\rm i} = \phi_{\rm i} \biggl( 1 + \frac{2}{5} R_{\nu}\biggr) + \frac{R_{\gamma}}{5}( 4 \sigma_{\rm B} - R_{\nu} \Omega_{\rm B}),\qquad \sigma_{\nu} = - \frac{R_{\gamma}}{R_{\nu}} \sigma_{\rm B} + 
\frac{k^2 \tau^2}{6 R_{\nu}} ( \psi_{\rm i} - \phi_{\rm i}).
\label{anis2}
\end{equation}
In the limit $\sigma_{\rm B}\to 0$ and $\Omega_{\rm B} \to 0$ 
this solution reproduces the standard adiabatic mode in the longitudinal gauge
(see third and fourth references in \cite{f7}). 
To follow the fate of  the magnetized adiabatic mode  through $\tau_{\rm eq}$ it is practical to exploit  the total density contrast on uniform 
curvature hypersurfaces (conventionally denoted by $\zeta$ \cite{f7,f8}) 
or the curvature perturbation on comoving orthogonal hypersurfaces (conventionally denoted by ${\cal R}$ \cite{f7,f8}) whose specific 
definitions, in terms of the variables of the longitudinal gauge, are
\begin{equation}
\zeta = - \psi - {\cal H} \frac{\delta \rho_{\rm t} + \delta\rho_{\rm B}}{\rho_{\rm t}'}, \qquad 
{\cal R} = - \psi - \frac{{\cal H}( {\cal H} \phi + \psi')}{{\cal H}^2- {\cal H}'},\qquad 
\zeta = {\cal R} + \frac{\nabla^2 \psi}{12\pi G a^2 (p_{\rm t} + \rho_{\rm t})}.
\label{Ham2}
\end{equation}
The first and second relations in Eq. (\ref{Ham2}) are the definitions 
of $\zeta$ and ${\cal R}$ in terms of the conformally Newtonian 
variables. The third relation in Eq. (\ref{Ham2}) can be obtained 
by substituting the definitions of $\zeta$ and ${\cal R}$ back into Eq. (\ref{Ham1}) 
and by recalling the background relations (\ref{FL}).
In the limit when the relevant wavelengths are all larger than the Hubble 
radius at the corresponding epoch, i.e. $k\tau < 1$, the third relation in (\ref{Ham2}) implies that ${\cal R} \simeq \zeta$.  The evolution 
equation for $\zeta$ can then be obtained by inserting the 
definition of $\zeta$ into Eq. (\ref{DC}); the result is 
\begin{equation}
\zeta' = - \frac{{\cal H}}{p_{\rm t} + \rho_{\rm t}} \delta p_{\rm nad} + 
\frac{{\cal H}}{p_{\rm t} + \rho_{\rm t}} \biggl( c_{\rm s}^2 - \frac{1}{3}\biggr) \delta\rho_{\rm B} - \frac{\theta_{\rm t}}{3}.
\label{zetaevol}
\end{equation}
The non-adiabatic pressure density variation 
 $\delta p_{\rm nad}$ can be written as a sum of the relative entropy fluctuations
over  the various components of the mixture
\begin{equation}
\delta p_{\rm nad}= \frac{1}{6 {\cal H} \rho_{\rm t}'} \sum_{{\rm i}\,{\rm j}} 
\rho_{\rm i}'\,\rho_{\rm j}' (c_{\rm s\,i}^2 -c_{\rm s\,j}^2) {\cal S}_{\rm i\,j},\qquad
{\cal S}_{i\rm j} = - 3 (\zeta_{\rm i} - \zeta_{\rm j}),\qquad c_{\rm s\,i}^2 = 
\frac{p_{\rm i}'}{\rho_{\rm i}'},
\label{defnad2}
\end{equation}
where ${\cal S}_{\rm i\,j}$ are the relative fluctuations in the entropy density 
that can be computed, from Eq. (\ref{defnad2}), in terms 
of the density contrasts of the individual fluids, i.e.   
\begin{eqnarray}
&& \zeta_{\rm c} = - \psi + \frac{\delta_{\rm c}}{3},\qquad 
\zeta_{\rm b} = - \psi + \frac{\delta_{\rm b}}{3},\qquad 
\zeta_{\nu} = - \psi + \frac{\delta_{\nu}}{4},\qquad
\zeta_{\gamma} = - \psi + \frac{\delta_{\gamma}}{4},
\nonumber\\
&& \zeta = \frac{\rho_{\nu}'  \zeta_{\nu} + \rho_{\gamma}' \zeta_{\gamma} + 
\rho_{\rm c}' \zeta_{\rm c} + \rho_{\rm b}' \zeta_{\rm b}}{\rho_{\rm t}'} + \zeta_{\rm B}
 ,\qquad \zeta_{\rm B} = \frac{\delta\rho_{\rm B}}{3( p_{\rm t} + \rho_{\rm t})}.
\end{eqnarray}
In the case of adiabatic (magnetized) initial conditions it can be 
easily verified that $\zeta_{\rm c} = \zeta_{\rm b} = \zeta_{\nu} = \zeta_{\gamma}$ 
so that $\delta p_{\rm nad}=0$.
Deep in the radiation-dominated epoch, for $\tau \ll \tau_{\rm eq}$, 
$c_{\rm s}^2 \to 1/3$ and, from Eq. (\ref{zetaevol}), $\zeta'=0$, so that 
\begin{equation}
\zeta = \zeta_{\rm i} \simeq {\cal R}_{\rm i},\qquad \zeta_{\rm i} = - \frac{3}{2}\phi_{\rm i}\biggl( 1 + \frac{4}{15} R_{\nu}\biggr) - \frac{R_{\gamma}}{5}( 4 \sigma_{\rm B} - R_{\nu} \Omega_{\rm B}).
\end{equation}
When the Universe becomes matter-dominated, after $\tau_{\rm eq}$,
$c_{\rm s}^2 \to 0$ and the second term at the right hand side of Eq. 
(\ref{zetaevol}) does contribute significantly at decoupling (recall that 
for $h^2 \Omega_{\rm matter} = 0.134$, $\tau_{\rm dec} = 2.36 \, \tau_{\rm eq}$).
Consequently, from Eq. (\ref{zetaevol}), recalling that $c_{\rm s}^2 = 
4 a_{\rm eq}/[ 3 ( 3 a + 4 a_{\rm eq})]$, we obtain 
\begin{equation}
 \zeta_{\rm f} = \zeta_{\rm i} - \frac{3 \,a\,R_{\gamma}\, \Omega_{\rm B}}{4 ( 3 a + 4 a_{\rm eq})},\qquad \Omega_{{\rm B}\,{\rm f}} = \Omega_{{\rm B}\,{\rm i}}. 
 \label{ZF}
\end{equation}
The inclusion of one (or more) adiabatic modes changes 
the form of Eq. (\ref{zetaevol}) and, consequently, the related solution 
(\ref{ZF}). For instance, in the case of the CDM-radiation non-adiabatic mode 
the relevant terms arising in the sum (\ref{defnad2}) are 
${\cal S}_{{\rm c}\gamma} = 
{\cal S}_{{\rm c}\nu} = {\cal S}_{\rm i}$ where ${\cal S}_{i}$ is the (constant) 
fluctuation in the relative entropy density initially present 
(i.e. for $\tau \ll \tau_{\rm eq}$). If  this is the case
 $\delta p_{\rm nad} =  c_{\rm s}^2 \rho_{\rm c} {\cal S}_{i}$ and Eq.
(\ref{zetaevol}) can be easily solved. The transfer matrix for magnetized CMB anisotropies can then be written as 
\begin{equation}
\pmatrix{
 \zeta_{{\rm f}} \cr
{\cal S}_{{\rm f}}\cr
\Omega_{{\rm B}\,{\rm f}}} = 
\pmatrix{{\cal M}_{\zeta \zeta} & {\cal M}_{\zeta{\cal S}} & 
{\cal M}_{\zeta {\rm B}}\cr
0 & {\cal M}_{{\cal S}{\cal S}}& {\cal M}_{{\cal S} {\rm B}}\cr
0 & 0& {\cal M}_{{\rm B} {\rm B}} }
\pmatrix{
 \zeta_{{\rm i}}\cr
{\cal S}_{{\rm i}}\cr
\Omega_{{\rm B}\,{\rm i}}}.
 \label{MAT1}
\end{equation}
In the case of a mixture of (magnetized) adiabatic and CDM-radiation
modes, we find, for $a > a_{\rm eq}$ 
\begin{equation}
{\cal M}_{\zeta \zeta} \to 1, \qquad {\cal M}_{\zeta {\cal S}} \to - \frac{1}{3},\qquad {\cal M}_{\zeta{\rm B}} - \frac{R_{\gamma}}{4},\qquad {\cal M}_{{\cal S}{\cal S}} \to 1,\qquad {\cal M}_{{\cal S}{\rm B}}\to 0,
\label{MAT2}
\end{equation}
and  ${\cal M}_{{\rm B}{\rm B}} \to 1$.
Equations (\ref{MAT1}) and (\ref{MAT2}) may be used, for instance, 
to obtain the magnetized curvature and entropy fluctuations 
at photon decoupling in terms of the same quantities evaluated 
for $\tau \ll \tau_{\rm eq}$. A full numerical analysis of the problem 
confirms the analytical results summarized by Eqs. (\ref{MAT1}) and 
(\ref{MAT2}). The most general initial condition for CMB anisotropies will then be a combination of (correlated) fluctuations receiving contribution 
from $\delta p_{\rm nad}$ and from the fully inhomogeneous 
magnetic field. To illustrate this point, the form of the Sachs-Wolfe (SW) plateau in the sudden decoupling limit will now be discussed. 
To compute the SW contribution we need 
to solve the evolution equation of the monopole of the temperature 
fluctuations in the tight coupling limit, i.e. from Eqs. (\ref{pb1}) and 
(\ref{pb2}),  
\begin{equation}
\delta_{\gamma}''+ \frac{{\cal H} R_{\rm b}}{ 1 + R_{\rm b}} \delta_{\gamma}' + 
\frac{k^2}{3} \frac{\delta_{\gamma}}{1 + R_{\rm b}} = 4 \psi'' + 
\frac{4 {\cal H} R_{\rm b}}{1 + R_{\rm b}} \psi' - \frac{4}{3} k^2 \phi - 
\frac{k^2}{3 ( 1 + R_{\rm b})} ( \Omega_{\rm B} - 4 \sigma_{\rm B}).
\label{monopole}
\end{equation}
In the sudden decoupling approximation the visibility function, i.e. ${\cal K}(\tau) = \kappa'(\tau) e^{-\kappa(\tau)}$  and the optical depth, i.e. $\epsilon^{-\kappa(\tau)}$ are approximated, respectively, by 
$\delta(\tau - \tau_{\rm dec})$ and by $\theta(\tau - \tau_{\rm dec})$. 
The power spectra of $\zeta$, ${\cal S}$ and $\Omega_{\rm B}$ are 
\begin{equation}
{\cal P}_{\zeta}(k) = {\cal A}_{\zeta} \biggl(\frac{k}{k_{\rm p}}\biggr)^{n_{r}-1},\qquad
{\cal P}_{{\cal S}}(k) 
 = {\cal A}_{\cal S} \biggl(\frac{k}{k_{\rm p}}\biggr)^{n_{s}-1},
\qquad 
{\cal P}_{\Omega}(k) = {\cal F}(\varepsilon) \overline{\Omega}_{{\rm B}\, L}^2 \biggl(\frac{k}{k_{L}}\biggr)^{2 \varepsilon},
\label{SP}
\end{equation}
where ${\cal A}_{\zeta}$, ${\cal A}_{{\cal S}}$ and $\overline{\Omega}_{{\rm B}\,L}$ 
are  constants and 
\begin{equation}
{\cal F}(\varepsilon) = \frac{4(6 - \varepsilon) ( 2 \pi)^{ 2 \varepsilon}}{\varepsilon ( 3 - 2 \varepsilon)
 \Gamma^2(\varepsilon/2)},\qquad  \overline{\Omega}_{{\rm B}\,\, L} = \frac{\rho_{{\rm B}\,L}}{\overline{\rho}_{\gamma}}, 
\qquad \rho_{{\rm B}\,\, L}=\frac{ B_{L}^2}{8\pi},\qquad \overline{\rho}_{\gamma}= a^{4}(\tau) \rho_{\gamma}(\tau).
\label{DEFB}
\end{equation}
To deduce Eqs. (\ref{SP}) and (\ref{DEFB}) the magnetic field has been regularized, according to a common practice \cite{f3,f4},  
over a typical comoving scale $L = 2\pi/k_{L}$  with a Gaussian window function and it has been assumed that the magnetic field intensity is stochastically distributed as 
\begin{equation}
\langle B_{i}(\vec{k},\tau) B^{j}(\vec{p},\tau) \rangle =  \frac{2\pi^2}{k^3} \,P_{i}^{j}(k)\, P_{\rm B}(k,\tau)\, \delta^{(3)}(\vec{k} + \vec{p}),
\label{Bcorr}
\end{equation}
where 
\begin{equation}
P_{i}^{j}(k) = \biggl(\delta_{i}^{j} - \frac{k_{i}k^{j}}{k^2} \biggr),\qquad P_{\rm B}(k,\tau) = A_{\rm B} 
\biggl(\frac{k}{k_{\rm p}}\biggr)^{\varepsilon}.
\label{MPS}
\end{equation}
As a consequence of Eq. (\ref{Bcorr}) the magnetic field does not break
the spatial isotropy of the background geometry.
The quantity $k_{\rm p}$ appearing in Eqs. (\ref{SP}) and (\ref{MPS}) is 
conventional pivot scale that is $0.05\, {\rm Mpc}$(see \cite{f6} for a discussion of other possible choices). Equations (\ref{SP}) and (\ref{DEFB}) hold 
for $0<\varepsilon < 1$. In this limit the magnetic energy spectrum is 
nearly scale-invariant. This means that the effect of the magnetic and thermal 
diffusivity scales (related, respectively, to the finite value of the 
conductivity and of the thermal diffusivity coefficient) do not affect 
the spectrum \cite{f4}. In the opposite limit, i.e. $\varepsilon \gg 1$ the value of the mode-coupling integral appearing in the two-point function of the magnetic energy density (and of the magnetic anisotropic stress) is dominated by ultra-violet effects related to the mentioned diffusivity scales \cite{f4}. Using then Eqs. (\ref{SP}) and (\ref{DEFB}) 
the $C_{\ell}$ can be computed for the region of the SW plateau (i.e.  for multipoles $\ell < 30$):
\begin{eqnarray}
&&C_{\ell} = \biggl[ \frac{{\cal A}_{\zeta}}{25} \,{\cal Z}_{1}(n_{r},\ell)  +
\frac{9}{100} \, R_{\gamma}^2  \overline{\Omega}^{2}_{{\rm B}\,L} {\cal Z}_{2}(\epsilon,\ell) - 
\frac{4}{25} \sqrt{{\cal A}_{\zeta} {\cal A}_{{\cal S}}} {\cal Z}_{1}(n_{r s},\ell) \cos{\gamma_{r s}}
\nonumber\\
&& +\frac{4}{25} \,{\cal A}_{{\cal S}} \,
{\cal Z}_{1}(n_{s},\ell)- 
\frac{3}{25} \sqrt{ {\cal A}_{\zeta}} \, R_{\gamma} \,\overline{\Omega}_{{\rm B}\, L}\,{\cal Z}_{3} (n_{r},\varepsilon, \ell) \cos{\gamma_{br}} 
\nonumber\\
&& + \frac{6}{25} \sqrt{ {\cal A}_{{\cal S}}}\,R_{\gamma} \overline{\Omega}_{{\rm B}\,L}\, {\cal Z}_{3}(n_{s},\varepsilon, \ell)\cos{\gamma_{b s}} \biggr],
\label{SWP}
\end{eqnarray}
where the functions ${\cal Z}_{1}$, ${\cal Z}_{2}$ and ${\cal Z}_{3}$ 
\begin{eqnarray}
{\cal Z}_{1}(n,\ell) &=& \frac{\pi^2}{4} \biggl(\frac{k_0}{k_{\rm p}}\biggl)^{n-1} 2^{n} \frac{\Gamma( 3 - n) \Gamma\biggl(\ell + 
\frac{ n -1}{2}\biggr)}{\Gamma^2\biggl( 2 - \frac{n}{2}\biggr) \Gamma\biggl( \ell + \frac{5}{2} - \frac{n}{2}\biggr)},
\label{Z1}\\
{\cal Z}_{2}(\varepsilon,\ell) &=& \frac{\pi^2}{2} 2^{2\varepsilon} {\cal F}(\varepsilon) \biggl( \frac{k_{0}}{k_{L}}\biggr)^{ 2 \varepsilon} \frac{ \Gamma( 2 - 2 \varepsilon) 
\Gamma(\ell + \varepsilon)}{\Gamma^2\biggl(\frac{3}{2} - \varepsilon\biggr) \Gamma(\ell + 2 -\varepsilon)},
\label{Z2}\\
{\cal Z}_{3}(n,\varepsilon,\ell) &=&\frac{\pi^2}{4} 2^{\varepsilon} 2^{\frac{n +1}{2}} \,\sqrt{{\cal F}(\varepsilon)}\, \biggl(\frac{k_{0}}{k_{L}}\biggr)^{\varepsilon} \biggl(\frac{k_{0}}{k_{\rm p}}\biggr)^{\frac{n + 1}{2}} \frac{ \Gamma\biggl(\frac{5}{2} - \varepsilon - \frac{n}{2}\biggr) \Gamma\biggl( \ell + 
\frac{\varepsilon}{2} + \frac{n}{4} - \frac{1}{4}\biggr)}{\Gamma^2\biggl(\frac{7}{4} - \frac{{\varepsilon}}{2} - \frac{n}{4}\biggr)
\Gamma\biggl( \frac{9}{4} + \ell - \frac{\varepsilon}{2} - \frac{n}{4} \biggr)},
\label{Z3}
\end{eqnarray}
are defined in terms of the magnetic tilt $\varepsilon$ and of a generic spectral index $n$ which may correspond, depending on the specific contribution, either to $n_{r}$ (adiabatic spectral index), or to $n_{s}$(non-adiabatic spectral index)  or even to $n_{rs} = (n_{r} + n_{s})/2$ (spectral index of the cross-correlation).  In Eq. (\ref{SWP}) $\gamma_{rs}$, $\gamma_{br}$ and $\gamma_{s b}$ are the correlation angles. In the absence of magnetic and non-adiabatic contributions and for Eqs. (\ref{SWP}) and Eq. (\ref{Z1}) imply that for $n_{r}=1$ (Harrison-Zeldovich spectrum) $ \ell (\ell +1) C_{\ell}/2\pi = 
{\cal A}_{\zeta}/25$ and WMAP data \cite{f5} would imply that 
${\cal A}_{\zeta}= 2.65 \times 10^{-9}$. Consider then the 
physical situation where on top of the adiabatic mode there is a magnetic contribution. If there is no correlation 
between the magnetized contribution and the adiabatic contribution, i.e. 
$\gamma_{b r} =\pi/2$, the SW plateau will be enhanced in comparison 
with the case when magnetic fields are absent.  The same situation 
arises when the two components are anti-correlated (i.e. $\cos{\gamma_{br}}<0$).
However, if the fluctuations are positively correlated (i.e.  
$\cos{\gamma_{b r}}>0$) the cross-correlation adds negatively to the 
sum of the two autocorrelations of $\zeta$ and $\Omega_{\rm B}$ so that 
the total result may be an  overall reduction of the power  with respect
to the case  $\gamma_{br} =\pi/2$. In Eq. (\ref{Z1}),(\ref{Z2}) and (\ref{Z3})
$k_{0} = \tau_{0}^{-1}$ where $\tau_{0}$ is the present observation time. 
Taking as typical parameters 
$h^2 \Omega_{\rm c} = 0.111$, $h^2 \Omega_{\rm b} = 0.023$ 
and $h^2 \Omega_{\rm r}= 4.15 \times 10^{-5}$ (with $h =0.73$),  the amplitude 
of the magnetic field intensity can be constrained by requiring 
that the adiabatic mode dominates. In the case when the magnetic and adiabatic 
contribution are totally anticorrelated (i.e. $\cos{\gamma_{b r}}= -1$), which 
is the most unfavourable case the bound on the protogalactic
field  read, at the present epoch and over a typical comoving scale 
$L = 2\pi/k_{L}$ with $k_{L} = {\rm Mpc}^{-1}$
\begin{equation}
B_{L} < 2.5 \times 10^{-9}\,\, {\rm G},\qquad n_{r} \simeq 0.951,\qquad \varepsilon \simeq 0.9.
\label{B1}
\end{equation}
As indicated, the bound (\ref{B1}) assumes a nearly scale-invariant (but slightly red \cite{f5}) adiabatic mode and the maximally allowed magnetic spectral tilt.
A further reduction of $\varepsilon$ leads to a slight relaxation of the bound; for 
instance for $\varepsilon \simeq 0.4$, $B_{L} < 6.3 \times 10^{-9}\,\, {\rm G}$.

In CMB physics is common practice to perform model-independent 
analysis on the parameter space of the allowed initial conditions 
by including, for instance, correlated (or anticorrelated) non-adiabatic modes in the game (see, for instance,  \cite{f6} and references therein). Up to now the effects related to fully inhomogeneous magnetic fields have been discussed within a different standard whose limitations were  
the impossibility of defining accurately both initial conditions and normalization of the magnetized (scalar) CMB anisotropies. The results reported here allow to overcome these difficulties and lead naturally to a strategy of parameter extraction where  large-scale
magnetic fields are treated consistently as a further degree of freedom in the space of the initial conditions.  According to this perspective, it will be important to pursue the analysis of small-scale effects by semi-analytical methods to corroborate and interpret more numerical 
studies related to parameter extraction. The first step in this direction would be to go to higher orders in the tight coupling expansion and generalize the standard treatment to the case when the pre-decoupling plasma is effectively magnetized.  Along these directions work is in progress.

\end{document}